\documentclass[12pt,a4]{article}
\usepackage{epsfig}

\newlength{\abstwidth}
\def\llgm{\left\lgroup\matrix}
\def\rrgm{\right\rgroup}

\def\gslash#1{\slash\hspace*{-0.20cm}#1}
\def\be{\begin{equation}} 
\def\ee{\end{equation}} 
\def\bq{\begin{equation}} 
\def\eq{\end{equation}} 
\def\bqa{\begin{eqnarray}} 
\def\eqa{\end{eqnarray}} 
\def\lsim{\roughly<}
\def\gsim{\roughly>}
\renewcommand{\theequation}{\arabic{section}.\arabic{equation}}

\begin{document}

\def\lsim{\mathrel{\rlap{\lower4pt\hbox{\hskip1pt$\sim$}}
    \raise1pt\hbox{$<$}}}         
\def\gsim{\mathrel{\rlap{\lower4pt\hbox{\hskip1pt$\sim$}}
    \raise1pt\hbox{$>$}}}         

\vspace{\fill}


\begin{center}
{\large \bf Two-body and Three-body Decays of Charginos in One-loop 
Order in MSSM}
\footnote{Dedicated to Dr. J. Kodaira, our friend and colleague, who passed away in Sept. 2006}
\\ 
\vspace{12 mm}

{
FUJIMOTO Junpei$~^{a}$, ISHIKAWA Tadashi$~^{a}$, \\
JIMBO Masato$~^{b}$,KON Tadashi$~^{c}$, \\
KURIHARA Yoshimasa$~^{a}$, KURODA Masaaki$~^{d}$
}
\vspace{4 mm}

{
$~^{a}~${\bf KEK,} {\it Oho, Tsukuba, Ibaraki 305-0801 Japan}\\
$~^{b}~${\bf Tokyo Management College,} {\it Ichikawa, Chiba 272-0001, Japan}\\
$~^{c}~${\bf Seikei University,} {\it Musashino, Tokyo 180-8633, Japan}\\
$~^{d}~${\bf Meiji Gakuin University,} {\it Totsuka, Yokohama 244-8539, Japan}\\
} 
\end{center}

\vspace{2 cm}
\noindent{\bf Abstract}\\
We present the renormalization scheme used in and the characteristic 
features of {\tt GRACE/SUSY-loop}, the package of the program
for the automatic calculation of the MSSM processes including one-loop order 
corrections.  The two-body and three-body decay widths of charginos 
in one-loop order evaluated by {\tt GRACE/SUSY-loop} are shown.


\section{Introduction}
The concept of supersymmetry (SUSY) \cite{susy, mssm} is considered the most  
promising extension of the Standard Model (SM) \cite{sm} of particle physics. 
Among the supersymmetric theories,
the minimal supersymmetric standard model (MSSM) \cite{mssm} is 
the most elaborated and well studied framework of SUSY.  
The existence of many new supersymmetric particles
in any SUSY model makes it very complicated and tedious to compute 
even a simple two-body decay 
width exactly. To overcome this problem, 
the Minami-tateya group of KEK has constructed a computational system 
{\tt GRACE/SUSY} \cite{graces1,graces2}
which automatically creates, for a given process,  all the Feynman diagrams and 
compute the Feynman amplitudes and subsequently the cross section or 
the decay width itself at tree level in the MSSM.

With the increase of more precise experimental data, in particular, 
in future colliders, it becomes apparent that  the inclusion of 
at least one-loop radiative  corrections is necessary for calculations of
cross sections of SUSY processes.
For this purpose, the SPA project has established the convention
of the SUSY parameters and loop calculations\cite{spa1}.
We have, therefore, extended {\tt GRACE/SUSY}
in order to incorporate radiative corrections in one-loop order
in the system.  The new system, called {\tt GRACE/SUSY-loop},
is constructed based on the same philosophy 
as for {\tt GRACE-loop} \cite{grace} developed by the Minami-tateya group,
which is the automatic computation system for the SM processes
including one-loop corrections. 
Some of the results computed by this system have already 
been presented in 
several workshops and symposiums \cite{res1, res2, res3}. 

In this paper we present the result of the thorough investigation of the
two-body and three-body decays of charginos in one-loop
order in the MSSM computed by the automatic computing system 
{\tt GRACE/SUSY-loop}. 
We also show the cross section of the chargino  pair production in $e^+e^-$ annihilation
and the subsequent decays  by combining the production cross section
and the decay rates.

The paper is organized as follows. In section two, the renormalization scheme 
used in {\tt GRACE/SUSY-loop} is explained. The features of {\tt the GRACE/SUSY-loop}
system are briefly given in section three. The numerical results
are presented in section four, and several comments on the numerical
results are given in section five.

\section{Renormalization scheme}
In this section we explain briefly the renormalization
scheme adopted in {\tt GRACE/SUSY-loop}.   Our approach is a straightforward
extension of the on-shell renormalization  in the SM used in {\tt GRACE}
\cite{grace}.  

The Lagrangian of the MSSM has been given in \cite{mk1, hikasa, rosiek}.
In terms of superfields, it is given as ( see \cite{mk1} for detail)
\bqa
   {\cal L} &=&~ \int d^2\theta {1\over 4} 
               [2 Tr({\bf W}{\bf W}) + WW + 2Tr({\bf W_s}{\bf W_s})] + h.c. 
               \nonumber \\
     & &+ \int d^2\theta d^2\bar\theta
          {\bf \Phi_\ell}^\dagger \exp[2(g{{\tau^a}\over 2}V^a
            + g^\prime {{Y_\ell}\over 2}V)] {\bf \Phi_\ell} \nonumber \\
     & &+ \int d^2\theta d^2\bar\theta \Phi_e^\dagger \exp(g^\prime Y_e V) 
            \Phi_e  \nonumber \\
     & &+ \int d^2\theta d^2\bar\theta 
           {\bf \Phi_q}^\dagger \exp[2(g{{\tau^a}\over 2}V^a+ 
             g^\prime {{Y_q}\over 2}V
          + g_s{{\lambda^\alpha}\over 2} V_s^\alpha)]{\bf \Phi_q} \nonumber \\
     & &+ \int d^2\theta d^2\bar\theta \Phi_u ^\dagger \exp(g^\prime Y_u V
          -g_s\lambda^{\alpha *}V_s^\alpha) \Phi_u  \nonumber  \\
     & &+ \int d^2\theta d^2\bar\theta \Phi_d^\dagger \exp(g^\prime Y_d V
          -g_s\lambda^{\alpha *}V_s^\alpha)\Phi_d  \nonumber \\
     & &+ \int d^2\theta d^2\bar\theta 
        {\bf \Phi_{H1}}^\dagger \exp[2(gT^aV^a+ g^\prime {{Y_{H1}}\over 2}V)] 
               {\bf \Phi_{H1}} \nonumber \\
     & &+ \int d^2\theta d^2\bar\theta 
          {\bf \Phi_{H2}}^\dagger\exp[2(gT^aV^a+ g^\prime {{Y_{H2}}\over 2}V)] 
               {\bf \Phi_{H2}} \nonumber \\
    &  &+{{\sqrt 2 m_e}\over v_1} \int d^2\theta {\bf \Phi_{H_1}}
               {\bf \Phi_\ell}\Phi_e + h.c.  \nonumber \\
    &  &-{{\sqrt 2 m_u}\over v_2} \int d^2\theta {\bf \Phi_{H_2}}
               {\bf\Phi_q}\Phi_u + h.c. \nonumber \\
    &  &+{{\sqrt 2 m_d}\over v_1} \int d^2\theta {\bf \Phi_{H_1}}
               {\bf\Phi_q}\Phi_d + h.c. \nonumber \\
    &  &-\mu \int d^2\theta  {\bf \Phi_{H1}}{\bf \Phi_{H2}}+ h.c.  \nonumber \\
    &  &+ {\cal L}_{soft}       \nonumber \\
    &  &+ {\cal L}_{gf}+{\cal L}_{ghost},  
\label{2.1}
\eqa
where ${\bf W}$, $W$, ${\bf W_s}$ are  superfield strengths corresponding 
to the SU(2)$_L$, U(1) and SU(3)$_c$ gauge-superfields, ${\bf V}$, $V$, ${\bf V_s}$,
respectively.
The component fields belonging to each superfield are given as follows
\bqa
    {\bf V} &=& (\vec W_\mu, \vec\lambda),  
              ~~~~~~~~~~~~~~~~SU(2)_L~{\rm gauge~boson}\nonumber \\
         V  &=& (B_\mu, \lambda),   ~~~~~~~~~~~~~~~~~U(1)~{\rm gauge~boson} \nonumber \\
    {\bf V_s} &=& (g_\mu^\alpha, \tilde g_\mu^\alpha), 
              ~~~~~~~~~~~~~~~~SU(3)_c~{\rm gauge~boson}\nonumber \\
    {\bf \Phi_{H_1}} &=& \Bigl( \llgm{{\bf H}_1 \cr {\bf \tilde H}_1}\rrgm\Bigr)
                      =\Bigl( \llgm{H^0_1 \cr H^-_1}\rrgm, ~~
                                \llgm{\tilde H^0_1 \cr \tilde H^-_1}\rrgm\Bigr), 
                            \nonumber \\
    {\bf \Phi_{H_2}} &=& \Bigl( \llgm{{\bf H}_2 \cr {\bf \tilde H}_2}\rrgm\Bigr)
                      =\Bigl( \llgm{H^+_2 \cr H^0_2}\rrgm, ~~
                                \llgm{\tilde H^+_2 \cr \tilde H^0_2}\rrgm\Bigr), 
             \nonumber \\
    {\Phi_\ell}&=& \Bigl(\llgm{ \nu_L \cr e_L}\rrgm, ~~\llgm{ \tilde\nu_L \cr \tilde e_L}\rrgm\Bigr), 
             \nonumber \\
    {\Phi_e}&=& (e_R, \tilde e_R^*), \nonumber \\
    {\Phi_q}&=& \Bigl(\llgm{ u_L \cr d_L}\rrgm, ~~\llgm{ \tilde u_L \cr \tilde d_L}\rrgm\Bigr), 
             \nonumber \\
    {\Phi_u}&=& (u_R, \tilde u_R^*), \nonumber \\
    {\Phi_d}&=& (d_R, \tilde d_R^*), 
\eqa
The SU(2)$_L$ doublet gauge-bosons and gauginos are denoted by
$\vec W_\mu^a$, $\vec\lambda^a$, while the singlet gauge-boson and
gaugino are denoted by $B_\mu$ and $\lambda$.  
The Higgs and higgsino doublets are denoted by ${\bf H}_i$ and 
$\tilde{\bf H}_i$ with $i=1,2$, respectively.
The soft SUSY breaking terms are expressed as
\bqa
   {\cal L}_{soft} &=&    -{1\over 2}M_1\lambda\lambda 
                        -{1\over 2}M_2\lambda^a\lambda^a 
                        -{1\over 2}M_3\tilde g^\alpha\tilde g^\alpha + h.c.
            \nonumber \\
            &&      -\tilde m_1^2{\bf H}_1^*{\bf H}_1 
                       -\tilde m_2^2{\bf H}_2^*{\bf H}_2
                       -( m_{12}^2{\bf H}_1{\bf H}_2 + h.c.)
                       - \sum _{\tilde f_m} \tilde m^2_{\tilde f_m}
                        \tilde f^*_m \tilde f_m \nonumber \\
     & &- {{\sqrt 2 m_u}\over v_2}A_u {\bf H_2}{\bf A}(q_L)A(u_R)
      + {{\sqrt 2 m_d}\over v_1}A_d {\bf H_1}{\bf A}(q_L)A(d_R)+h.c.\nonumber \\
     & &+ {{\sqrt 2 m_e}\over v_1}A_e {\bf H_1}{\bf A}(\ell_L)A(e_R)
       + h.c.  \label{2.3}
\eqa
where the sum in the sfermion mass terms runs for 
$\tilde f_m = \tilde f_L$ and $\tilde f_R$.
The sign convention of our $A_f$ is opposite to the convention used by others. 

Sfermion mass eigenstates
are denoted by $\tilde f_i$ with $i=1,2$ which are 
the mixture of the left-handed ($\tilde f_L$)  and the right-handed 
sfermions ($\tilde f_R$).  The sfermion mass matrix is diagonalized as,
\bqa   
    \llgm{\cos\theta_f & \sin\theta_f \cr -\sin\theta_f & \cos\theta_f}\rrgm
    \llgm{m^2_{\tilde f_L} & m^2_{\tilde f_{LR}} \cr 
          m^{2*}_{\tilde f_{LR}} & m^2_{\tilde f_R} }\rrgm
    \llgm{\cos\theta_f & -\sin\theta_f \cr \sin\theta_f & \cos\theta_f}\rrgm \nonumber
\eqa
\bqa   
  = \llgm{m^2_{\tilde f_1} & 0 \cr 0 & m^2_{\tilde f_2} }\rrgm ,
\label{2.4}
\eqa
where
\bqa
    m^2_{\tilde f_L} &=&~ \tilde m^2_{\tilde f_L}+ m^2_f 
                        +M_Z^2\cos 2\beta(T_{3f}-Q_f s^2_W) , \nonumber \\
    m^2_{\tilde f_R} &=&~ \tilde m^2_{\tilde f_R}+ m^2_f 
                        +M_Z^2\cos2\beta Q_fs^2_W,\label{2.5} \\
    m^2_{\tilde f_{LR}}&=& \cases{-m_u(\mu \cot\beta+A_u), & $f=u$\cr
                                 -m_f(\mu \tan\beta+A_f), & $f=d,e$ }
                    ~.   \nonumber
\eqa
Their masses and the mixing angles satisfy
the  relations originating from the SU(2)$_L$ conditions on their left-handed
soft SUSY-breaking mass terms,
 $\tilde m^2_{\tilde u_L}=\tilde m^2_{\tilde d_L}$,
 $\tilde m^2_{\tilde e_L}=\tilde m^2_{\tilde \nu_e}$ etc.
For the third generation of sfermions, for example,  
\bqa
   \cos^2\theta_t m^2_{\tilde t_1} + \sin^2\theta_t m^2_{\tilde t_2} 
   -m_t^2&=& 
   \cos^2\theta_b m^2_{\tilde b_1} + \sin^2\theta_b m^2_{\tilde b_2} 
   -m_b^2+M_W^2 \cos 2\beta,  \nonumber \\
   m^2_{\tilde \nu\tau}
   &=& \cos^2\theta_\tau m^2_{\tilde \tau_1} 
     +\sin^2\theta_\tau m^2_{\tilde \tau_2} -m_\tau^2+ M_W^2 \cos2\beta.
      \nonumber \\
   && \label{2.6}
\eqa

Regarding now the Lagrangian (\ref{2.1}) and (\ref{2.3}) as
a bare Lagrangian and renormailzing all the quantities,
we separate the renormalized Lagrangian and
its counterterms.
The renormalization constants are introduced as follows.\\
\noindent [Standard Model sector]\\
\bqa
  {\rm gauge~ bosons}~~~~~~~~~
     \vec{W}_{\mu 0} &=& Z_W^{1/2} \vec{W}_\mu,     \nonumber \\
           B_{\mu 0} &=& Z_B^{1/2} B_\mu,           \nonumber \\
           g_{\mu 0} &=& Z_{\rm gluon}^{1/2} g_{\mu},   \nonumber \\
  {\rm gauge~couplings}~~~~~~~~~~~
          g_0 &=& Z_gZ_W^{-3/2} g,                  \nonumber \\
   g^\prime_0 &=& Z_{g^\prime}Z_B^{-3/2} g^\prime,  \nonumber \\
       g_{s0} &=& Z_{g_s}Z_{\rm gluon}^{-3/2} g_s,      \nonumber \\
  {\rm fermions}~~~~~~~~~~~~~~~~
    \Psi_{fL_0}&=& Z_f^{L~1/2}\Psi_{fL},~~~~~~f=u,d,\cdots,\nu_e,e,\cdots ~,
                                                   \nonumber \\
    \Psi_{fR_0}&=& Z_f^{R~1/2}\Psi_{fR},~~~~~~f=u,d,\cdots,e,\cdots ~,
                                                   \nonumber \\
    m_{f_0} &=& m_f + \delta m_f,~~~~~~f=u,d,\cdots,e,\cdots ~.
\label{2.7}
\eqa
\noindent [SUSY sector]\\
\bqa
   {\rm Higgs~ bosons}~~~~~~~~~
      {\bf H}_{i0} &=& Z_{H_i}^{1/2} {\bf H}_i ,~~~~~~~~~~~~~~~~~~~~ 
           i = 1,2 ~, \nonumber \\ 
        v_{i0} &=& Z^{1/2}_{H_i} (v_i-\delta v_i) , ~~~~~~~~~~~
            i=1,2 ~, \nonumber \\ 
      m_{i0}^2 &=& Z_{H_i}^{-1} (m_i^2+\delta m_i^2) , ~~~~~~~~~
            i=1,2 ~, \nonumber \\ 
      (m_{12}^2)_0 &=& Z_{H_1}^{-1/2} Z_{H_2}^{-1/2} 
                    (m_{12}^2+\delta m_{12}^2) ,  \nonumber \\ 
   {\rm sfermions}~~~~~~~~~~
    \llgm{\tilde f_1 \cr \tilde f_2}\rrgm_0 &= & 
   \llgm{Z_{\tilde f_1 \tilde f_1}^{1/2} & Z_{\tilde f_1 \tilde f_2}^{1/2}\cr
         Z_{\tilde f_2 \tilde f_1}^{1/2} & Z_{\tilde f_2 \tilde f_2}^{1/2}}
                 \rrgm
    \llgm{\tilde f_1 \cr \tilde f_2}\rrgm , \nonumber \\
            && ~~~~~~~~~~~~~~~~~~~~~~~~~ f=u,d,\cdots, e,\cdots ~, \nonumber \\
    (\tilde \nu_i)_0 &=& Z^{1/2}_{\tilde \nu_i} \tilde \nu_i , 
                 ~~~~~~~~~~~~~~~~ i=e,\mu,\tau ~, \nonumber \\ 
    (m_{\tilde f_i}^2)_0 &=&  m_{\tilde f_i}^2 + \delta m_{\tilde f_i}^2 ,
                            ~~~~~~~~~ f=u,d,\cdots, e,\cdots ~, \nonumber \\
            && ~~~~~~~~~~~~~~~~~~~~~~~~~~ i=1,2 ~, \nonumber \\
    (m^2_{\tilde \nu_i})_0 &=& m^2_{\tilde \nu_i}+\delta m^2_{\tilde \nu_i},
                  ~~~~~~~~~~ i = e,\mu,\tau ~, \nonumber \\ 
    (\theta_f)_0 &=& \theta_f + \delta \theta_f, ~~~~~~~~~~~~~ f=u,d,\cdots, e,\cdots ~, \nonumber \\ 
    {\rm Inos}~~~~~~~~~~~~~~~~~~~~~
   \vec \lambda_0 &=& Z^{1/2}_{\lambda^w} \vec \lambda , \nonumber \\ 
        \lambda_0 &=& Z^{1/2}_\lambda  \lambda , \nonumber \\ 
   \tilde{\bf H}_{i0} &=& Z^{1/2} _{\tilde H_i} \tilde{\bf H}_i ,
         ~~~~~~~~~~~~~~~i=1,2 ~, \nonumber \\ 
   \tilde g^\alpha_0 &=& Z_{\tilde g}^{1/2} \tilde g^\alpha ,   \nonumber \\       
         \mu_0 &=& \mu + \delta\mu ,   \nonumber \\ 
         M_{10}&=& M_1 + \delta M_1 ,  \nonumber \\ 
         M_{20}&=& M_2 + \delta M_2 ,  \nonumber \\ 
         M_{30}&=& M_3 + \delta M_3 , ~~~~~
              (m_{\tilde g 0}=m_{\tilde g}+ \delta m_{\tilde g}) ~, ~ 
\label{2.8}
\eqa
where in the Higgs boson part,
\bq
     m_i^2 = \tilde m_i^2+|\mu|^2 ~, ~~ i=1,2 ~.   \label{2.9}
\eq

     Summing up, we have ($3+7N_G$) wavefunction renormalization constants,
($3+3N_G$) mass counterterms in the non-SUSY sector,
and ($7+13N_G$) wavefunction renormalization constants and ($9+10N_G$)
mass, vacuum-expectation-value and mixing-angle counterterms in the SUSY 
sector.

Several comments are in order.
The relation (\ref{2.4}) $\sim$ (\ref{2.6}) are originally 
satisfied by bare quantities, but they are also valid among
the renormalized quantities, although the renormalized masses
are not necessarily equal to the pole masses.
In a similar way, (\ref{2.9}) can be understood as the relation among 
the renormalized quantities. Note that at this state all the renormalized
parameters in the Lagrangian are simply parameters of the model,
and only upon imposing the renormalization conditions to be specified 
in this section, they are 
related to the physical quantities.

In the Higgs and higgsino sectors, the wavefunction renormalization
constants are introduced to each unmixed bare doublet state.  The mixing
angles in the Higgs, chargino and neutralino sectors are defined as
the angles which diagonalize the renormalized mass matrices.
Therefore, there 
appear no bare mixing angles or counterterms for the mixing angles 
of charginos and neutralinos in our scheme.  See \cite{chankowski1}.

In place of $\delta m_1^2$, $\delta m_2^2$ and $\delta m_{12}^2$,
we use the mass counterterm of the CP odd Higgs particle, $\delta M_A^2$,
and two counterterms of the tadpole interactions, $\delta {\cal T}_1$
and $\delta {\cal T}_2$, which are given by the linear combination of
$\delta m_1^2$, $\delta m_2^2$ and $\delta m_{12}^2$;
\bqa
     \delta M^2_A &=&  
          s_\beta^2 \delta m_1^2 + c_\beta^2 \delta m_2^2
        - 2c_\beta s_\beta \delta m_{12}^2  \\
     && -{{M_Z^2}\over 2}(c_\beta^2-s_\beta^2)^2 
          [\delta Z_Z + \delta Z_{H_1} + \delta Z_{H_2}
          -{{2c_\beta^2}\over{c_\beta^2-s_\beta^2}}{{\delta v_1}\over {v_1}}
          +{{2s_\beta^2}\over{c_\beta^2-s_\beta^2}}{{\delta v_2}\over {v_2}}] , 
             \nonumber \\
    \delta {\cal T}_1 &=& v_1\delta m_1^2-m_1^2\delta v_1 
                         +v_2\delta m_{12}^2-m_{12}^2\delta v_2 \nonumber \\
            & & +{1\over 8} (\delta g^2 + \delta g^{\prime 2})
                        (v_1^2-v_2^2)v_1   
               +{1\over 8} (g^2 + g^{\prime 2})
               \{(v_2^2-3v_1^2)\delta v_1 + 2v_1v_2\delta v_2\}\nonumber \\
            & &+ {1\over 8} (g^2 + g^{\prime 2})
                \{(2v_1^3-v_1v_2^2)\delta Z_{H_1} 
                  -v_1v_2^2\delta Z_{H_2} \} , \\
    \delta {\cal T}_2 &=& (1 \leftrightarrow 2 ) ~{\rm in}~ \delta {\cal T}_1 .
\eqa

We introduce the gauge-fixing Lagrangian in terms of the
renormalized fields as we have done in {\tt GRACE-loop} for SM \cite{grace}.
No renormalization constants are introduced for gauge fixing constants
and no counterterm Lagrangian corresponding to the gauge-fixing 
Lagrangian appears.  The renormalization of the ghost fields is not
necessary in one-loop order.

\vspace{0.5 cm}

We use the on-shell renormalization scheme.
The renormalization conditions employed in {\tt GRACE/SUSY-loop} are 
the following set of conditions.  Using these conditions, we can express
all the renormalization constants and counterterms in terms of 
the linear combination of the two-point functions
evaluated at some specific renormalization points. 

\noindent $\underline{\rm gauge~sector}$ \\
We use the on-mass-shell condition for $W$, $Z$ and photon.
In addition, we require that the residue of the photon
propagator at the pole position is one, and that 
the $A_\mu$-$Z_\mu$
transition vanishes for the on-shell photon.
 
\noindent $\underline{\rm SM~ fermions}$\\
We use the same renormalization conditions adopted  in {\tt GRACE-loop}.
Namely, we require the on-mass-shell condition  and the residue condition
that the residue of the fermion propagator at the pole is one.

\noindent $\underline{\rm Higgs~sector}$\\
We impose the on-mass-shell and the residue condition for CP odd
Higgs, $A^0$, and the decoupling of $Z_\mu$ and $A^0$ on the
mass-shell of $A^0$.
For the CP even Higgs, we impose the on-mass-shell condition
for the heavier Higgs, $H^0$.
The above three conditions together with one of the conditions imposed
in the gauge sector determine four renormalization constants
$\delta H_1$, $\delta H_2$, ${{\delta v_1}\over{v_1}}$
and ${{\delta v_2}\over{v_2}}$.

Note that we have not adopted  the often erroneously used renormalization
condition,
${{\delta v_1}\over{v_1}}= {{\delta v_2}\over{v_2}}$, since
this condition violates the gauge invariance \cite{yamada}. 

\noindent $\underline{\rm tadpole ~terms}$\\
Identical to {\tt GRACE-loop}, we require that the tadpole terms in the
renormalized Lagrangian vanish by itself and the tadpole
counterterms cancel the one-loop tadpole contributions.

\noindent $\underline{\rm chargino ~sector}$\\
We impose the on-mass-shell condition on both
$\tilde\chi_1^+$ and $\tilde\chi_2^+$.
In addition, we impose the residue condition on
$\tilde\chi_1^+$.

\noindent $\underline{\rm neutralino~sector}$\\
We impose the on-mass-shell and the residue condition
on the lightest neutralino, $\tilde\chi_1^0$.

\noindent $\underline{\rm sfermion ~sector}$\\
We impose the on-mass-shell condition and the residue condition
on all the seven sfermions in each generation.
In addition, we impose that there is no induced mixing between
physical $\tilde f_1$ and $\tilde f_2$.
The counterterm for the slepton mixing angle is determined by the
SU(2)$_L$ relation (\ref{2.6}) upon  introducing the renormalization 
constants (\ref{2.7}) and (\ref{2.8}).   In the squark sector,
there are two counterterms $\delta \theta_u$ and $\delta \theta_d$ 
for each generation.  We fix $\delta\theta_u$ \cite{guasch}
by 
\bq
    \delta\theta_u = {1\over 2}{{\Sigma_{\tilde u_1\tilde u_2}(m_{\tilde u_1}^2)
                     +\Sigma_{\tilde u_1\tilde u_2}(m_{\tilde u_2}^2) }\over
                     {m_{\tilde u_2}^2-m_{\tilde u_1}^2}} ,
\label{2.13}
\eq
while $\delta\theta_d$ is fixed by the SU(2)$_L$ relation (\ref{2.6})
upon introducing the renormalization constants (\ref{2.7}) and (\ref{2.8}).

\noindent $\underline{\rm charge~renormalization}$\\
The charge (electromagnetic coupling constant) is defined as in the standard model
\cite{nlg}.

\noindent $\underline{\rm QCD ~sector}$\\
    We impose the on-mass-shell condition and the residue condition
for gluons and gluinos.  The counterterm $\delta Z_{g_s}$ is determined by 
the minimal subtraction with dimensional reduction ($\overline{\rm DR}$).

\vspace{0.5 cm}

The explicit expression of the renormalization constants
 in terms of two-point functions is given in Appendix A.

A couple of comments are worthwhile at this stage.
In our renormalization scheme, the pole mass of $h^0$, $H^\pm$, $\tilde\chi^0_2$,
$\tilde\chi^0_3$ and $\tilde\chi^0_4$  is different from its Born value.
Therefore, the mixing angles of the Higgs sector and the neutralino sector are not
directly related to the pole masses.  They are considered as effective parameters.
This means, in particular, that a quantity "$\tan\beta$" which includes 
one-loop corrections is not defined and not used in {\tt GRACE/SUSY-loop}.
The relation between the parameter $\tan\beta$ used in {\tt GRACE/SUSY-loop}
and
the experimentally observed "$\tan\beta$" which includes higher order
corrections depends on how "$\tan\beta$" is actually defined.

In addition to the wavefunction renormalization constants introduced in 
(\ref{2.7}) and (\ref{2.8}), we need to introduce the ultraviolet finite external 
wavefunction renormalization 
constant $\delta Z^{ext}$ for each particle for which the residue condition
is not imposed on its propagator, namely for $W^\pm$, $Z^0$, $H^0$, 
$h^0$, $H^\pm$, $\tilde\chi^\pm_2$, $\tilde\chi_2^0$,  $\tilde\chi_3^0$,  
$\tilde\chi_4^0$. The expression of $\delta Z^{ext}$ in terms of 
two-point functions is given in Appendix B.

Since the coupling constant of 
the soft SUSY-breaking Yukawa interaction 
among Higgs and sfermions, $A_f$, appears always in combination with $m_f$,
we use in {\tt GRACE/SUSY-loop} $m_fA_f$ and its counterterm $\delta(m_fA_f)$
as independent variables, which helps avoid partly the numerical 
instability of the counterterm at large $\tan\beta$.  Explicitly
\bqa
   \delta(m_fA_f) &=&{1\over 2}(\delta m^2_{\tilde f_2}
                     -\delta m^2_{\tilde f_1})\sin 2\theta_f
                     +\delta\theta_f(m^2_{\tilde f_2}-m^2_{\tilde f_1})
                      \cos2\theta_f \nonumber \\
                  & &-\cases{\delta (m_t\mu\cot\beta) &$f=u,c,t$ \cr
                             \delta (m_f\mu\tan\beta) &$t=d,s,b,e,\mu,\tau$} ,
\label{2.14}
\eqa
since according to (\ref{2.4}) $A_f$ is related to the sfermion masses 
$m_{\tilde f_1}^2$ and  $m_{\tilde f_2}^2$ as
\bq
   m_fA_f = \cos\theta_f\sin\theta_f(m_{\tilde f_2}^2-m_{\tilde f_1}^2)
         -\cases{m_f \mu \cot\beta & $f=u,c,t$ \cr
                 m_f \mu \tan\beta & $f=d,s,b,e,\mu,\tau$} .
\label{2.15}
\eq

The system can easily accommodate
 different renormalization schemes by re-expressing  the renormalization
constants and the mass counterterms  in terms of
different linear combinations of the two-point functions.\\ 

A brief comparison of our scheme with other earlier studies is worthwhile
to clarify the difference and also the possible scheme dependence
in different schemes.
There are plenty of papers on the renormalization schemes of the MSSM,
and it is beyond our scope to compare all of them.  
The earlier works on the MSSM renormalization (see for example,
\cite{ayamada,chankowski1, dabelstein1}) are naturally concerned with
radiative corrections in the Higgs sector. 
Later, the study of the renormalization of the other sector,
gaugino and higgsino sector and sfermion sector followed ( see for example,
\cite{ pierce, hagiwara, hollik1, guasch2}).
In \cite{ayamada, dabelstein1,  chankowski1, hollik1, fritzsche} the on-shell
renormalization is used, while the so-called $\overline{DR}$
is used in \cite{hagiwara}.  Since the number of the particles exceeds the
number of the parameters appearing in MSSM, one cannot impose the 
on-shell conditions on all the particles.  Therefore, there can be  many 
variants even among the on-shell scheme, depending on which particles
are put on-mass-shell.

The renormalization scheme we adopt in {\tt GRACE/SUSY-loop} is close
to the scheme given by \cite{chankowski1}, but there
are a couple of differences.   
In \cite{chankowski1}, different from our scheme,  
the gauge-fixing Lagrangian is introduced 
in terms of bare fields.  Therefore, in their scheme, they
need extra renormalization conditions to fix the
renormalization constants for gauge-parameters.  
Another difference of our scheme from the others lies in 
the renormalization condition imposed on $\delta v_i$.  
The condition used in \cite{dabelstein1, chankowski1, hollik1} $\delta v_1/v_1
= \delta v_2/v_2$ is not preferable on the ground of
potentially violating the Ward identity.  We use the on-mass-shell
condition for $H^0$ in place of the condition on the
vacuum expectation value, which leads to our expression of $\delta\tan\beta$ 
which is also different from others.

\setcounter{equation}{0}
\section{Features of {\tt GRACE/SUSY-loop}}
We present in this section several features of the current 
version of the system {\tt GRACE/SUSY-loop}.  
Some of them  have been given in \cite{res1}.

In order to check and detect possible errors in the system 
we have used  the technique of the non-linear gauge (NLG),
by adding the SUSY interactions to the gauge fixing functions
of the SM \cite{nlg}.  Explicitly, we used the following 
gauge fixing functions\footnote{In principle, we can also add non-linear
sfermion interactions.  We have not attempted this extension,
since due to the mixing the resultant Lagrangian becomes too
lengthy and cumbersome.}
\bqa
   {\cal L}_{\rm gf} &=&  -{1\over{\xi_W}}\vert F_{W^+} \vert^2 
                      -{1\over{2\xi_Z}}(F_Z)^2
                      -{1\over{2\xi_\gamma}}(F_\gamma)^2 , \\
   F_{W^\pm} &=& (\partial_\mu \pm ie\tilde\alpha A_\mu  
                  \pm ig c_W\tilde\beta Z_\mu) W^{\pm\mu} \nonumber \\
         & &   \pm i\xi_W {g\over 2}(v + \tilde\delta_H H^0 + \tilde\delta_h h^0 
              \pm i\tilde\kappa G^0)G^\pm ,
            \\
   F_Z &=& \partial_\mu Z^\mu+\xi_Z{{g_Z}\over 2}(v +\tilde\epsilon_H H^0
                +\tilde\epsilon_h h^0)G^0 , 
              \\
   F_\gamma &=& \partial_\mu A^\mu, 
\eqa
where
\bq
    v = \sqrt{v_1^2+v_2^2} ~~.
\eq
The gauge fixing functions contain now seven free parameters,
\bq
 \tilde\alpha,~~~, \tilde\beta,~~~ \tilde\delta_H,~~~ \tilde\delta_h,~~~
 \tilde\kappa,~~~  \tilde\epsilon_H,~~~\tilde\epsilon_h. 
\eq
Since physical results are independent on the NLG parameters,
they must vanish in the sum of the Feynman amplitudes for physical
processes.  The test using the NLG parameters provides us with a more
powerful tool for the check of the system than the test using
the linear gauge parameters, because each NLG parameter is concerned
with many kinds of amplitudes which are not in the same gauge-independent
sub-set in the linear gauge \cite{res3}.

In the actual computation of decay widths and production cross sections
in {\tt GRACE/SUSY-loop}, we use the 't Hooft-Feynman gauge with $\xi_V=1$.
For the consistency check of the computation, we use
the vanishing of the ultraviolet divergences and the infrared singularities
in the sum of the loop and
the soft photon/gluon radiation diagrams,
as well as the stability in the sum of the soft photon/gluon
radiation and the hard photon/gluon radiation diagrams
against the change of the photon/gluon energy cut-off.

The input parameters of {\tt GRACE/SUSY-loop} are
\bqa
     && e, g_s,M_W,M_Z,M_{A^0},\tan\beta, \mu,M_1,M_2,M_3 ,  \nonumber\\
     && m_u,m_d,m_e,\cdots , \nonumber\\
     &&  m_{\tilde u_1}, m_{\tilde u_2},
         m_{\tilde d_1}, m_{\tilde d_2}, 
         m_{\tilde e_1}, m_{\tilde e_2}, m_{\tilde \nu_e},  \cdots ,
         \nonumber \\
     && \theta_u,\theta_d,\theta_e, \cdots ,
\label{3.1}
\eqa
Using (\ref{2.6}), we fix the remaining two masses of the sfermions
in each generation.  Note that the width of the unstable particles
is neglected  in the computation of the amplitudes.

The coupling constants $A_f$  of the soft SUSY-breaking Yukawa interaction 
among Higgs and sfermions are not our independent 
input parameters, since  they are 
 expressed in terms of sfermion masses and the mixing angles
as (\ref{2.15}).
We don't use the GUT relations $M_1={5\over 3}\tan^2\theta_W M_2$,
$M_3={{g_s^2}\over {e^2}}\sin^2\theta_W M_2$ at our energy scale.

We have chosen the input values of the parameters in such a way that
they reproduce the the SPA1 pole mass values \cite{spa1} as close
as possible.
Since the input values of the SPA1 parameters \cite{spa1} are defined by 
the $\overline{DR}$ scheme, it is not possible to reproduce exactly 
the same values as those proposed by SPA1.
Note that the lightest neutralino ($\tilde\chi_1^0$) is 
the lightest SUSY-particle (LSP) both in SPA1 and in our parameter choice.

In the numerical calculation we adopt two numerical sets (A) and (B).  
which are given in Table 1 and 2, respectively.
For the fermion masses of the SM and gauge boson masses,
we used the following values (in unit of GeV) both for  sets (A) and (B), 
\bqa
  && M_W=80.35, ~~~~~ M_Z=91.1876, \nonumber \\
  && m_e=0.51099906\times 10^{-3}, ~~m_\mu = 105.658389\times 10^{-3},
     ~~m_\tau=1.7771,\nonumber \\
  && m_{\nu_e}= m_{\nu_\mu}=  m_{\nu_\tau}=0,\\
  && m_u=58.0\times 10^{-3}, ~~m_c=1.5, ~~m_t=178.0, \nonumber \\
  && m_d=58.0\times 10^{-3}, ~~m_s=92.0\times 10^{-3}, ~~m_b = 4.7,
  \nonumber 
\eqa
while for the strong coupling constant, we used
\bq
     \alpha_s = 0.12 ~~.
\eq
The (one-loop improved) mass of Higgs particles, charginos and neutralinos 
in the set (A) and (B)  is given in Table 3 and 4, respectively.

\begin{table}
\footnotesize{
\begin{tabular}{|c| c| c| c| c| c| c|}
\hline
   $\tan\beta$ & $\mu$ & $M_1$ & $M_2$ & $M_3$ & $M_{A^0}$  \\
\hline
    10.00 & 399.31 & 100.12 & 197.52 & 610 & 424.9    \\  
\hline
\end{tabular}

\begin{tabular}{|c| c| c| c| c| c| c| c|c|c|}
\hline
    $m_{\tilde u_1}$ & $m_{\tilde u_2}$ & $m_{\tilde d_1}$ & $m_{\tilde d_2}$ & 
    $m_{\tilde e_1}$ & $m_{\tilde e_2}$ & $m_{\tilde \nu_e}$ &
    $\cos\theta_u$  &  $\cos\theta_d$  &  $\cos\theta_e$  \\
\hline
 545.67 & 563.44 & 545.50 & 569.03 & 125.50 & 190.14 &172.70 &  
 2.4$\times 10^{-3}$  &  0.011  &  1.1$\times 10^{-4}$\\
\hline
\hline
    $m_{\tilde c_1}$ & $m_{\tilde c_2}$ & $m_{\tilde s_1}$ & $m_{\tilde s_2}$ & 
    $m_{\tilde \mu_1}$ & $m_{\tilde \mu_2}$ & $m_{\tilde \nu_\mu}$  &
    $\cos\theta_c$  &  $\cos\theta_s$  &  $\cos\theta_\mu$  \\ 
\hline
     545.66 & 563.45 & 545.52 & 568.97 & 125.43 & 190.16 & 172.69 &  
     0.063  &  0.018  & 0.023 \\
\hline
\hline
$m_{\tilde t_1}$ & $m_{\tilde t_2}$ & $m_{\tilde b_1}$ & $m_{\tilde b_2}$ & 
    $m_{\tilde \tau_1}$ & $m_{\tilde \tau_2}$ & $m_{\tilde \nu_\tau}$ & 
    $\cos\theta_t$  &  $\cos\theta_b$  &  $\cos\theta_\tau$  \\
\hline
    368.53 & 583.79 & 450.12 & 544.38 & 107.71 & 195.08 & 170.63&
    0.722  & 0.967  &  0.314 \\
\hline
\end{tabular}
}
\caption{The value of the MSSM input parameters for set (A)}
\end{table}       

\begin{table}
\footnotesize{
\begin{tabular}{|c| c| c| c| c| c| c|}
\hline
   $\tan\beta$ & $\mu$ & $M_1$ & $M_2$ & $M_3$ & $M_{A^0}$  \\
\hline
    10.00 & 399.15 & 100.13 & 157.53 & 610 & 431    \\  
\hline
\end{tabular}

\begin{tabular}{|c| c| c| c| c| c| c| c|c|c|}
\hline
    $m_{\tilde u_1}$ & $m_{\tilde u_2}$ & $m_{\tilde d_1}$ & $m_{\tilde d_2}$ & 
    $m_{\tilde e_1}$ & $m_{\tilde e_2}$ & $m_{\tilde \nu_e}$ &
    $\cos\theta_u$  &  $\cos\theta_d$  &  $\cos\theta_e$  \\
\hline
 506.48 & 524.14 & 506.07 & 530.14 & 163.22 & 187.37 &169.64 &
   9.4$\times 10^{-5}$  &  8.5$\times 10^{-4}$  &  9.1$\times 10^{-5}$\\
\hline
\hline
    $m_{\tilde c_1}$ & $m_{\tilde c_2}$ & $m_{\tilde s_1}$ & $m_{\tilde s_2}$ & 
    $m_{\tilde \mu_1}$ & $m_{\tilde \mu_2}$ & $m_{\tilde \nu_\mu}$  &
    $\cos\theta_c$  &  $\cos\theta_s$  &  $\cos\theta_\mu$  \\ 
\hline
     506.47 & 524.16 & 506.07 & 530.14 & 163.19 & 187.38 & 169.64 &
     0.033  &  1.6$\times 10^{-5}$  & 0.019 \\
\hline
\hline
$m_{\tilde t_1}$ & $m_{\tilde t_2}$ & $m_{\tilde b_1}$ & $m_{\tilde b_2}$ & 
    $m_{\tilde \tau_1}$ & $m_{\tilde \tau_2}$ & $m_{\tilde \nu_\tau}$ & 
    $\cos\theta_t$  &  $\cos\theta_b$  &  $\cos\theta_\tau$  \\
\hline
    345.37 & 556.78 & 469.43 & 507.15 & 150.07 & 190.39 & 170.02&
   0.5567  & 0.9266  &  0.271 \\
\hline
\end{tabular}
}
\caption{The value of the MSSM input parameters for set (B)}
\end{table}

\begin{table}
\begin{tabular} {|c|c|c|c|}
\hline
     $h^0$  & $H^0$ &  $A^0$ & $H^\pm$  \\
\hline
       107.12  &  425.30  &  424.90 &  432.75 \\
\hline   
\end{tabular}

\begin{tabular}{| c | c| c| c| c|c|}
\hline 
  $\tilde{\chi}^+_1$ & $\tilde{\chi}^+_2$ &  $\tilde{\chi}^0_1$ & $\tilde{\chi}^0_2$ &
  $\tilde{\chi}^0_3$ & $\tilde{\chi}^0_4$ \\
\hline
  184.2 & 421.2  & 97.75 & 184.62 & 398.30 & 413.39 \\
\hline
\end{tabular}
\caption{The pole mass of Higgs, charginos and neutralinos in set (A)}
\end{table}

\begin{table}
\begin{tabular} {|c|c|c|c|}
\hline
     $h^0$  & $H^0$ &  $A^0$ & $H^\pm$  \\
\hline
       122.50  &  431.40  &  431.0  &  438.73 \\
\hline   
\end{tabular}

\begin{tabular}{| c | c| c| c| c|c|}
\hline 
  $\tilde{\chi}^+_1$ & $\tilde{\chi}^+_2$ &  $\tilde{\chi}^0_1$ &
  $\tilde{\chi}^0_2$ & $\tilde{\chi}^0_3$ & $\tilde{\chi}^0_4$ \\
\hline
  147.08 & 418.8  & 97.61 & 147.4 & 404.0 & 418.8 \\
\hline
\end{tabular}
\caption{The pole mass of Higgs, charginos and neutralinos in set (B)}
\end{table}

The set (A) is almost the same as the SPA1a$^\prime$ parameter set, 
in which not only the heavier chargino $\tilde{\chi}^+_2$ but 
also the lighter chargino 
$\tilde{\chi}^+_1$ can decay into various two bodies 
because the lighter chargino is heavier than some sleptons. 
For the set (B), on the other hand, the mass of the lighter 
chargino $m_{\tilde{\chi}^+_1}$ 
is smaller than the mass of all sfermions $m_{\tilde{f}}$ as well as the sum 
$m_W + m_{\tilde{\chi}^0_1}$. 
This means that the lighter chargino $\tilde{\chi}^+_1$ cannot 
decay into any two bodies and 
$\tilde{\chi}^+_1$ has only  three body decay modes, $f\bar{f}\tilde{\chi}^0_1$.

\setcounter{equation}{0}
\section{Numerical results}
The one-loop electroweak corrections on various two-body decay widths of 
the lighter $\tilde{\chi}^+_1$ and heavier chargino 
$\tilde{\chi}^+_2$ for the parameter set (A) 
are shown in Table 5 and 6, respectively, where 
we do not display the decay modes with small branching 
fraction Br$< 0.1\%$. 
While all decays in Table 5 are the electroweak processes, the process 
$\tilde{\chi}^+_2 \to \bar{b} \tilde{t}_1$ 
in Table 6 gets both the electroweak and QCD corrections
through the loop contributions and the photon/gluon emissions. 
We define $\Gamma \equiv \Gamma_0 + \delta\Gamma$, where $\Gamma_0$ and $\delta\Gamma$ are 
the improved Born decay width and the one-loop correction, respectively. 
Note that the improved Born decay width is different from the Born width.
We obtain $\Gamma_0$ by replacing the tree-level masses
by the one-loop renormalized pole-masses presented
in Tables 3 and 4 in the tree amplitudes.

\begin{table*}[htb]
\newcommand{\m}{\hphantom{$-$}}
\newcommand{\cc}[1]{\multicolumn{1}{c}{#1}}
\renewcommand{\arraystretch}{1.2} 
\begin{tabular}{@{}llllll}
\hline
&   & $\Gamma_0$ (GeV) & $\Gamma$ (GeV) & $\delta\Gamma / \Gamma_0$ & Br
 \\
\hline
 & $\tilde{\chi}^+_1 \to \nu_\tau \tilde{\tau}^+_1$  &
   $3.91\times 10^{-2}$ & $3.78\times 10^{-2}$ & $-3.3\%$ & $50.11\%$ \\
 & $\tilde{\chi}^+_1 \to \nu_\mu \tilde{\mu}^+_1$    & 
   $1.33\times 10^{-4}$ & $1.19\times 10^{-4}$ & $-10.2\%$ & $0.16\%$ \\
 & $\tilde{\chi}^+_1 \to \tau^{+} \tilde{\nu}_\tau$  &
   $1.47\times 10^{-2}$ & $1.48\times 10^{-2}$ & $+0.1\%$ & $19.58\%$ \\
 & $\tilde{\chi}^+_1 \to \mu^{+} \tilde{\nu}_\mu$    &
   $1.06\times 10^{-2}$ & $1.07\times 10^{-2}$ & $+1.0\%$ & $14.24\%$ \\
 & $\tilde{\chi}^+_1 \to e^{+} \tilde{\nu}_e$    &
   $1.06\times 10^{-2}$ & $1.07\times 10^{-2}$ & $+1.0\%$ & $14.22\%$ \\
 & $\tilde{\chi}^+_1 \to W^+ \tilde{\chi}^0_1$   &
   $9.65\times 10^{-4}$ & $1.28\times 10^{-3}$ & $+32.3\%$ & $1.69\%$ \\
\hline
\end{tabular}\\[2pt]
\caption{One-loop corrections on $\tilde{\chi}^+_1$ decay widths for set (A)}
\end{table*}
\begin{table*}[htb]
\newcommand{\m}{\hphantom{$-$}}
\newcommand{\cc}[1]{\multicolumn{1}{c}{#1}}
\renewcommand{\arraystretch}{1.2} 
\begin{tabular}{@{}llllll}
\hline
&   & $\Gamma_0$ (GeV) & $\Gamma$ (GeV) & $\delta\Gamma / \Gamma_0$ & Br
 \\
\hline
 & $\tilde{\chi}^+_2 \to \nu_\tau \tilde{\tau}^+_2$  &
   $1.54\times 10^{-1}$ & $1.48\times 10^{-1}$ & $-3.9\%$ & $4.20\%$ \\
 & $\tilde{\chi}^+_2 \to \nu_\mu \tilde{\mu}^+_2$    &
   $1.36\times 10^{-1}$ & $1.46\times 10^{-1}$ & $+7.5\%$ & $4.13\%$ \\
 & $\tilde{\chi}^+_2 \to \nu_e \tilde{e}^+_2$    &
   $1.36\times 10^{-1}$ & $1.46\times 10^{-1}$ & $+7.6\%$ & $4.14\%$ \\
 & $\tilde{\chi}^+_2 \to \tau^{+} \tilde{\nu}_\tau$  &
   $6.89\times 10^{-2}$ & $5.70\times 10^{-2}$ & $-17.3\%$ & $1.61\%$ \\
 & $\tilde{\chi}^+_2 \to \mu^{+} \tilde{\nu}_\mu$    &
   $4.33\times 10^{-2}$ & $5.38\times 10^{-2}$ & $+24.2\%$ & $1.52\%$ \\
 & $\tilde{\chi}^+_2 \to e^{+} \tilde{\nu}_e$    &
   $4.32\times 10^{-2}$ & $5.37\times 10^{-2}$ & $+24.4\%$ & $1.52\%$ \\
 & $\tilde{\chi}^+_2 \to W^+ \tilde{\chi}^0_1$ &
   $1.93\times 10^{-1}$ & $2.07\times 10^{-1}$ & $+7.0\%$ & $5.85\%$ \\
 & $\tilde{\chi}^+_2 \to W^+ \tilde{\chi}^0_2$    &
   $8.66\times 10^{-1}$ & $9.93\times 10^{-1}$ & $+14.6\%$ & $28.12\%$ \\
 & $\tilde{\chi}^+_2 \to Z \tilde{\chi}^+_1$    &
   $7.53\times 10^{-1}$ & $8.56\times 10^{-1}$ & $+13.7\%$ & $24.26\%$ \\
 & $\tilde{\chi}^+_2 \to h^0 \tilde{\chi}^+_1$  &
   $5.97\times 10^{-1}$ & $6.07\times 10^{-1}$ & $+1.75\%$ & $17.20\%$ \\
 & $\tilde{\chi}^+_2 \to \bar{b} \tilde{t}_1$   &
   $2.82\times 10^{-1}$ & $2.57\times 10^{-1}$ & 
$\left\{ \begin{array}{l}-8.9\%({\rm ELWK})\\+1.8\%({\rm QCD}) \end{array}\right.$& $7.43\%$ \\
\hline
\end{tabular}\\[2pt]
\caption{One-loop corrections on $\tilde{\chi}^+_2$ decay widths for set (A)}
\end{table*}

In Table 7, we show the one-loop electroweak and QCD corrections 
to the various 3-body decay widths of 
$\tilde{\chi}^+_1$ in the parameter set (B), for which the 
two-body decays of $\tilde{\chi}^+_1$ are kinematically forbidden. 
The three-body decay widths of $\tilde{\chi}^+_2$ are not shown,
since, decaying dominantly into two bodies, it has extremely small 
three-body decay branching ratios.
Note that for the decay modes involving quarks 
$\tilde{\chi}^+_1 \to q\bar{q}\tilde{\chi}^0_1$, 
the electroweak and the QCD corrections are separately given. 

\begin{table*}[htb]
\newcommand{\m}{\hphantom{$-$}}
\newcommand{\cc}[1]{\multicolumn{1}{c}{#1}}
\renewcommand{\arraystretch}{1.2} 
\begin{tabular}{@{}llllll}
\hline
&   & $\Gamma_0$ (GeV) & $\Gamma$ (GeV) & $\delta\Gamma / \Gamma_0$ & Br
 \\
\hline
 & $\tilde{\chi}^+_1 \to e^+ \nu_e \tilde{\chi}^0_1$    
 & $4.42\times 10^{-6}$ & $4.84\times 10^{-6}$ & $+9.4\%$ & $20.18\%$ \\
 & $\tilde{\chi}^+_1 \to \mu^+ \nu_\mu \tilde{\chi}^0_1$    
 & $4.42\times 10^{-6}$ & $4.84\times 10^{-6}$ & $+9.4\%$ & $20.18\%$ \\
 & $\tilde{\chi}^+_1 \to \tau^+ \nu_\tau \tilde{\chi}^0_1$ 
 & $6.46\times 10^{-6}$ & $7.22\times 10^{-6}$ & $+11.8\%$ & $30.09\%$ \\
 & $\tilde{\chi}^+_1 \to u \bar{d} \tilde{\chi}^0_1$   
 & $3.35\times 10^{-6}$ & $3.55\times 10^{-6}$ & 
 $\left\{ \begin{array}{l}-0.2\%({\rm ELWK})\\+6.3\%({\rm QCD}) \end{array}\right.$
& $14.81\%$ \\
 & $\tilde{\chi}^+_1 \to c \bar{s} \tilde{\chi}^0_1$  
 & $3.33\times 10^{-6}$ & $3.54\times 10^{-6}$ & 
 $\left\{ \begin{array}{l}-0.2\%({\rm ELWK})\\+6.3\%({\rm QCD}) \end{array}\right.$
& $14.74\%$ \\
\hline
\end{tabular}\\[2pt]
\caption{One-loop corrections on $\tilde{\chi}^+_1$ decay widths for set (B)}
\end{table*}

We should note that the lighter chargino $\tilde{\chi}^+_1$ is the next  
lightest SUSY-particle (NLSP) in the set (B). 
Since NLSP must be first produced by accelerators,
it is important to study the production processes 
and the experimental signals of the chargino $\tilde{\chi}^+_1$. 
We calculate the full-one-loop electroweak correction for the chargino 
pair production $e^+e^- \to \tilde{\chi}^+_1 \tilde{\chi}^-_1$ 
at the future linear colliders ( See also \cite{eechi1lp, kilian, kilian2}). 
In Fig.1 the energy dependence of the total cross section is shown for 
the lighter chargino pair production at $e^+e^-$ colliders for the set (B) (and set (A)).  
The one-loop electroweak correction is of order of $-10\%$.
We note that the radiative correction of this order can be detectable at the proposed linear collider.  

\begin{figure}[htb]
\epsfig{figure=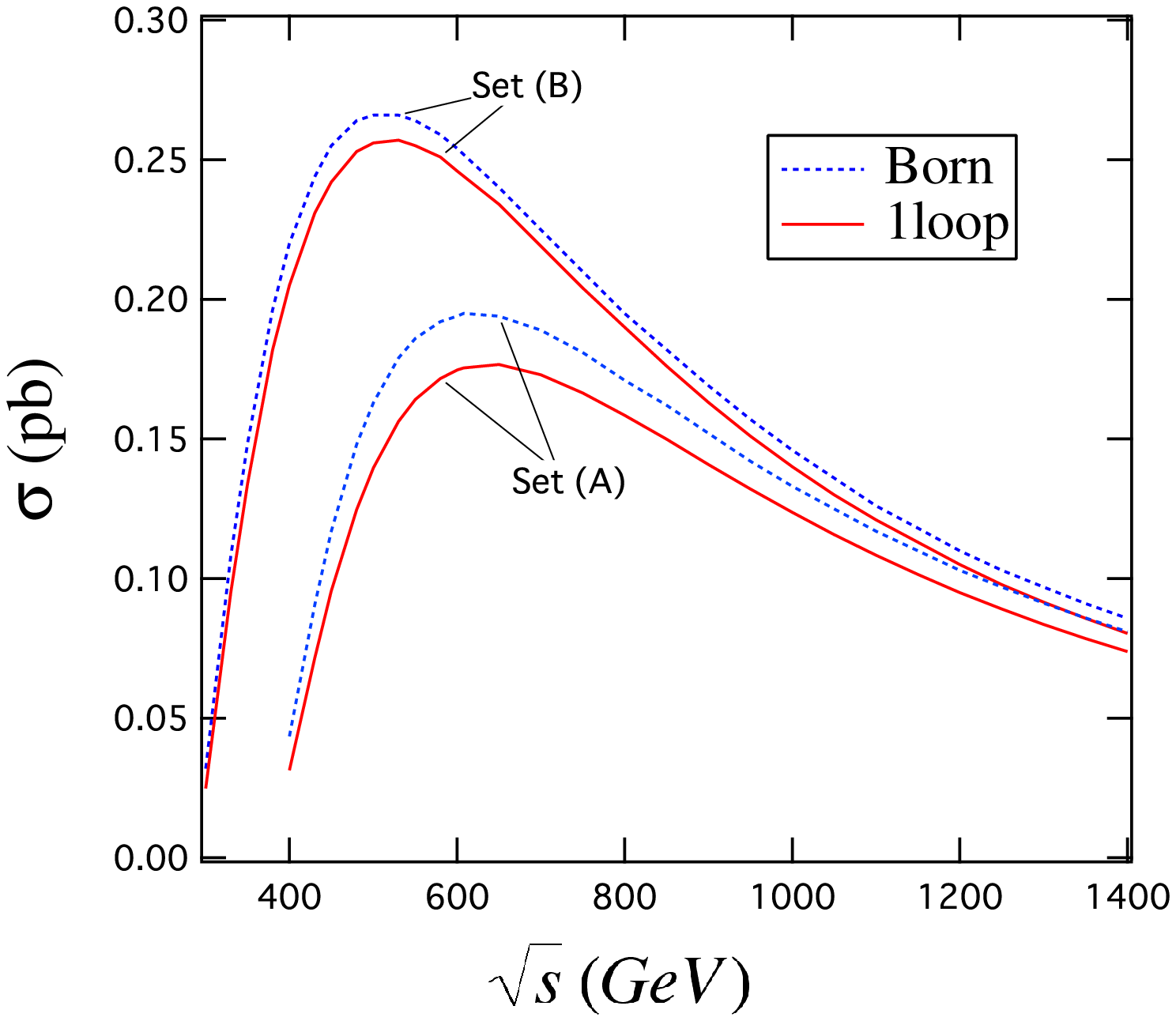,height=10cm,width=10cm,angle=0}
\caption{Total cross section for chargino pair production 
$e^+e^- \to \tilde{\chi}^+_1 \tilde{\chi}^-_1$ for the set (A) and set (B). 
Solid line and dotted line denotes $\sigma_{1loop}$ and $\sigma_{\rm BORN}$, respectively.}
\label{fig1}
\end{figure}

By combining the production cross section (Fig.1) and the 
decay branching ratios (Table 7) for the set (B), 
we obtain the one-loop corrected cross sections 
for the direct experimental signals. 
Fig.2 shows the energy dependence of the cross section for 
the two types of the chargino signals, 
$e^+e^- +$ missing energies and 4-jets $+$ missing energies,
at $e^+e^-$ colliders for the set (B).

\begin{figure}[htb]
\epsfig{figure=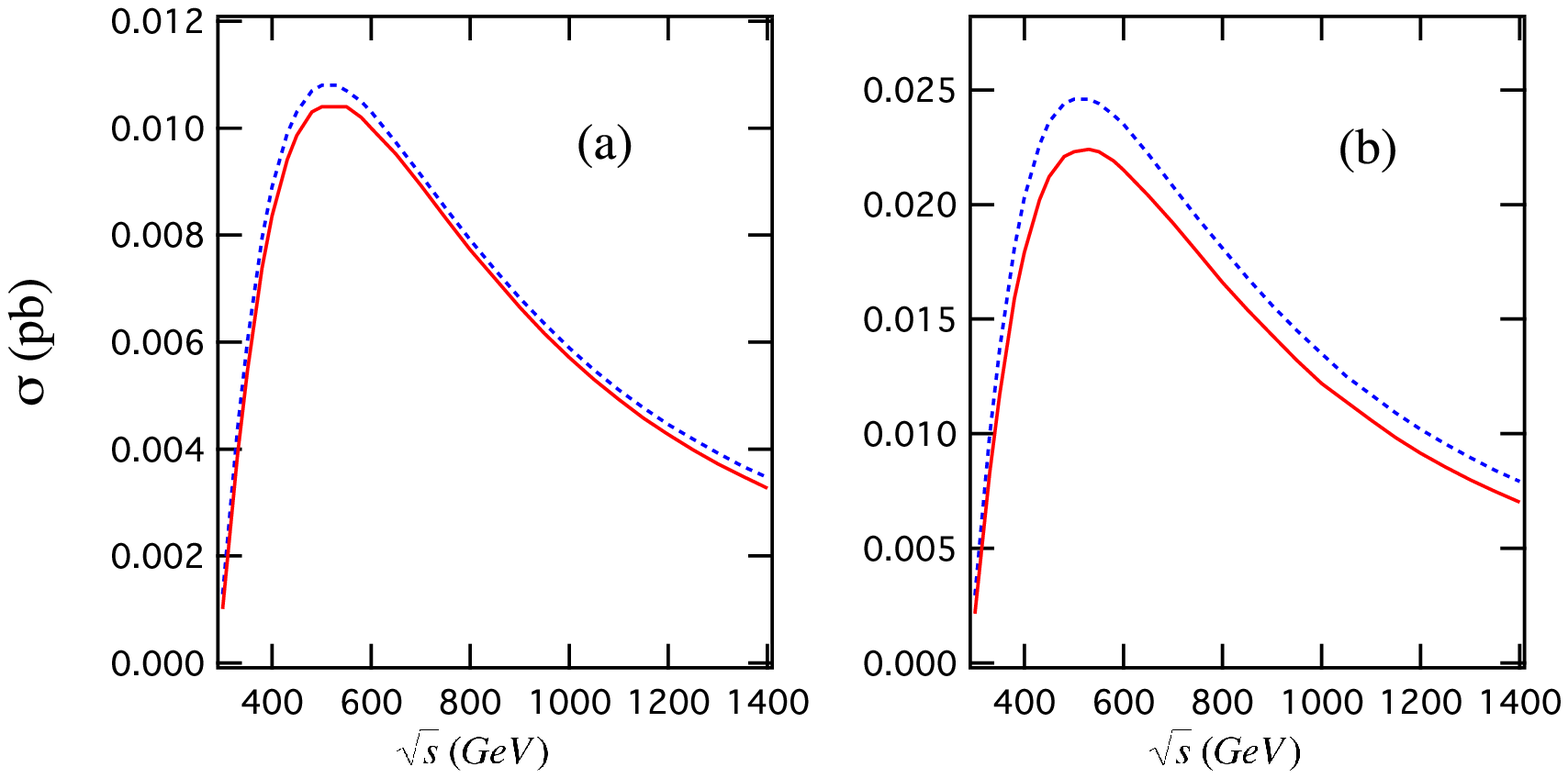,height=8cm,width=14.3cm,angle=0}
\caption{Total cross section for 
$e^+e^- \to \tilde{\chi}^+_1 \tilde{\chi}^-_1 \to 
(e^+ \nu_e \tilde{\chi}^0_1) (e^- \bar{\nu}_e \tilde{\chi}^0_1)$ (a) and 
$e^+e^- \to \tilde{\chi}^+_1 \tilde{\chi}^-_1 \to 
(q \bar{q'} \tilde{\chi}^0_1) (q \bar{q'} \tilde{\chi}^0_1)$ (b). 
Solid line and dotted line denotes $\sigma_{1loop}$ and $\sigma_{\rm BORN}$, respectively.}
\label{fig2}
\end{figure}

\setcounter{equation}{0}
\section{Comments}
  The total cross section is a sum of the one-loop corrected cross section
and the cross section of the hard photon radiation.  The former is negative
and large, while the latter is positive and large, originating mainly from
the initial radiation.  In our calculation for the hard photon radiation,
we do not set an energy cut in the upper limit nor any angle cut.

In the calculation of the one-loop correction of the decay widths the situation 
is almost the same except that in the evaluation of the decay widths the contribution of the real photon (gluon) emission from both initial and final states is important. 
The correction proportional to the fermion mass $m_f$ is expected to be large 
for the third generation $\tau$, $t$ and $b$, and 
an additional enhancement can emerge in the case of large $\tan\beta$ for 
$\tau$ and $b$. 
Note that we set $\tan\beta=10$ for both sets (A) and (B). 
We see these effects in the numerical results presented in Tables 5, 6 and 7, 
where the branching ratios of the $\tau$ modes are larger than $e$ and $\mu$ modes. 

The radiative correction on the chargino pair production in the SPA1a$^\prime$ scenario (set (A)) 
has been studied by the Wien group \cite{eechi1lp}. 
Unfortunately, as they adopt different treatment of the photon emission correction, 
the direct comparison of our result for the full electroweak correction 
with theirs is not possible. 
For the comparison we extract the "weak" correction 
$\Delta\sigma_{\rm weak}$
defined by  
\begin{eqnarray}
\Delta\sigma_{\rm weak} &\equiv& \sigma_{\rm elwk}
 - \sigma_{\rm BORN}*\delta_{\rm QED} -\sigma^{\rm initial}_{\rm hard}.
\end{eqnarray}
We find that the result of {\tt GRACE/SUSY-loop} is consistent with the previous result \cite{eechi1lp}.

In the parameter setting (A), the produced chargino dominantly decays
into two bodies, because the mass of the chargino $m_{\tilde{\chi}^+_1}$ ($=184.2$GeV) 
is larger than the mass of the lighter charged sleptons, $m_{\tilde{\ell}_1}$,
and $m_{\tilde{\nu}_\ell}$ 
as well as $m_W + m_{\tilde{\chi}^0_1}$ (see Tables 1 and 3). 
Since the chargino cannot decay into any squark in two-body decay modes and 
BR($\tilde{\chi}^+_1$ $\to$ $W^+ \tilde{\chi}^0_1$ $\to$ $qq' \tilde{\chi}^0_1$) 
is negligibly small (see Table 5), we cannot expect the signals with the quark-jets 
from the $\tilde\chi_1^+$ pair production. 
We find also that 
BR($\tilde{\ell}_1$ $\to$ $\ell \tilde{\chi}^0_1$) $=1$ and 
BR($\tilde{\nu}_\ell$ $\to$ ${\nu}_\ell \tilde{\chi}^0_1$) $=1$ in the parameter setting (A). 
This means that the most plausible experimental signal of the chargino pair production is 
the lepton pair plus the large missing energies. 
The precise measurement of the energy and the momentum of the $\tau$ leptons is particularly 
important because $\tau^+ \tau^-$ signal is the dominant mode in this case. 

In the parameter setting (B), the chargino $\tilde\chi_1^+$ with 
$m_{\tilde\chi_1^+}= 147.0$GeV is lighter than in set (A) in which 
$m_{\tilde\chi_1^+}= 184.2$GeV.
We find the apparent shift of the production peak to the lower energy due to 
the different chargino mass value used in the set (A) and (B). 
As has been discussed in the previous section, 
two-body decays of $\tilde{\chi}^+_1$ are kinematically forbidden in the set (B). 
We can use both types of signals, the lepton pair plus large missing energies and
the quark-jets plus large missing energies,
for the chargino detection. 

We have developed a tool for the full automatic one-loop calculation of the MSSM processes, 
{\tt GRACE/SUSY-loop}, which is characterized by the gauge symmetric and the on-shell renormalization scheme, 
and includes the various self-consistency check schemes. 
It is certainly a useful tool for the present and future precise analyses of 
the experimental data.

\vspace{1 cm}
{\bf Acknowledgement}\\
We would like to thank Y. Shimizu, T. Kaneko, G. B\'elanger and F. Boudjema
for fruitful discussions.  This work was partly supported by 
Japan Society for the Promotion of Science
under Grant-in-Aid for Scientific Research (B) (No.17340085).

\vfill\eject

\renewcommand{\theequation}{\Alph{section}.\arabic{equation}}
\setcounter{section}{1}
\setcounter{equation}{0}
\section*{Appendix A.  Expressions of conterterms}
In this Appendix, we list the expression of the counterterms and the
renormalization constants in terms of two-point functions.

The wavefunction renormalization constants are 
expanded in the one-loop order as
\bqa
    Z_X^{1/2} &=& 1+{1\over 2}\delta Z_X, \\
    Z_{XY}^{1/2} &=& 
     \cases{ 1+{1\over 2}\delta Z_{XY}, & $X=Y$ \cr
               {1\over 2}\delta Z_{XY}, & $X \ne Y$}
\eqa
We use the following abbreviations;
\bqa
  &&c_W = \cos\theta_W,~  c_L=\cos\phi_L,~ c_R =\cos\phi_R,~ 
  c_\alpha=\cos\alpha,~ c_\beta=\cos\beta, \nonumber \\
  &&s_W = \sin\theta_W,~  s_L=\sin\phi_L,~ s_R =\sin\phi_R,~ 
  s_\alpha=\sin\alpha,~ s_\beta=\sin\beta,
\eqa
where $\phi_L$ and $\phi_R$ are the mixing angles which diagonalize
the chargino mass matrix (see (2.10) of \cite{mk1}), and $\alpha$ 
is the mixing angle in the CP even Higgs sector.

\noindent $\underline{\rm gauge~sector}$\par
\bqa
  \delta Z_W = & Re [ \Pi_{AA}^\prime (0) 
         -2{{c_w}\over{s_W}} {{\Pi_{AZ}(0)}\over{M_Z^2}} 
         + {{\Pi_W(M_W^2)-c_W^2\Pi_{ZZ}(M_Z^2)}\over{s_W^2M_Z^2}}~] ,\\
  \delta Z_B = & Re [\Pi_{AA}^\prime (0) 
         +2{{s_W}\over{c_W}} {{\Pi_{AZ}(0)}\over{M_Z^2}} 
         - {{\Pi_W(M_W^2)-c_W^2\Pi_{ZZ}(M_Z^2)}\over{c_W^2M_Z^2}}~] ,\\
  \delta Z_g = & Re [\Pi_{AA}^\prime (0) 
         -{{2c_W^2+1}\over{c_Ws_W}} {{\Pi_{AZ}(0)}\over{M_Z^2}} 
         + {{\Pi_W(M_W^2)-c_W^2\Pi_{ZZ}(M_Z^2)}\over{s_W^2M_Z^2}}~] .
\eqa
Other counterterms appearing in the gauge sector are expressed
in terms of $\delta Z_W$, $\delta Z_B$ and $\delta Z_g$.  For example,
\bqa
  {{\delta g}\over g} &=& \delta Z_g-{3\over 2}\delta Z_W , \\
  {{\delta g^\prime}\over{g^\prime}}&=& -{1\over 2}\delta Z_B , \\
  {{\delta s_W}\over{s_W}}&=& -c_W^2\bigl(
      {{\delta g}\over g}-{{\delta g^\prime}\over{g^\prime}}\bigr) , \\
  {{\delta c_W}\over{c_W}}&=& +s_W^2\bigl(
      {{\delta g}\over g}-{{\delta g^\prime}\over{g^\prime}}\bigr) .
\eqa

\noindent $\underline{\rm fermion~sector}$\par
\bqa
      \delta m_f &=& -Re\Sigma_f^S(m_f) -m_f Re\Sigma_f^V(m_f) , \\
   \delta Z_f^R  &=& Re\Sigma_f^V(m_f)+ Re\Sigma_f^A(m_f) \nonumber \\
                 && ~~   + 2m_f[~Re\Sigma_f^{S\prime}(m_f)
                     + m_f Re\Sigma_f^{V\prime}(m_f)~] , \\
   \delta Z_f^L  &=& Re\Sigma_f^V(m_f)- Re\Sigma_f^A(m_f) \nonumber \\
                 && ~~    + 2m_f[~Re\Sigma_f^{S\prime}(m_f)
                     + m_f Re\Sigma_f^{V\prime}(m_f)~] .
\eqa
where the selfenergy function of Dirac 
fermion $f$ is decomposed as
\bq
    \Sigma_f(\gslash q) \equiv \Sigma_f^S(q^2) {\bf 1} + 
                              \Sigma_f^P(q^2) \gamma_5
     + \Sigma_f^V(q^2)\gslash q +\Sigma_f^A(q^2)\gslash q \gamma_5. 
\label{a.14}
\eq

\noindent $\underline{\rm tadpole~ terms}$\par
\bq
    \delta {\cal T}_i = T^{loop}_{\phi_i^0} , ~~~~~~~~~~~   i=1,2 ~,
\eq
where $\delta {\cal T}_i$ is the tadpole counterterms.

\noindent $\underline{\rm Higgs~sector}$\par
\bqa
  \delta M_A^2 &=&-Re\Sigma_{A^0A^0}(M_A^2)
     + M_A^2Re \Sigma_{A^0A^0}^\prime(M_A^2) ,\\
 \llgm{ \delta Z_{H_1} \cr  \delta Z_{H_2} \cr
       {{\delta v_1}\over{v_1}} \cr {{\delta v_2}\over{v_2}}}\rrgm
 &=& \llgm{ 2c_\beta^2 &  & c_\beta^2\cr
          -2s_\beta^2 &  & -s_\beta^2 \cr
          c_\beta^2 & C(i,j) & +{1\over 2}\cr
          -s_\beta^2  &   & -{1\over 2}}\rrgm
  \llgm{ {1\over{c_\beta s_\beta}}
         \bigl({{Re\Sigma_{A^0Z}(M^2_A)}\over{M_Z}} \bigr) \cr
         C~Re\Sigma^\prime_{A^0A^0}(M_A^2) \cr
         C\delta X \cr
         Cc_{2\beta} \delta Y}\rrgm ,
\eqa
where
\bqa
     C &=& {2\over{\sin 2\beta(\sin 2\beta+\sin 2\alpha)}}, 
            \nonumber \\
  C(1,2) &=& 3s_\beta^2c_\beta^2-s_\alpha^2c_\beta^2
            +2s_\alpha c_\alpha s_\beta c_\beta, \nonumber \\
  C(1,3) &=&c_\beta^2(c_\alpha^2-s_\beta^2),
          \nonumber \\
  C(2,2) &=& 3s_\beta^2c_\beta^2-c_\alpha^2s_\beta^2
           +2s_\alpha c_\alpha s_\beta c_\beta, \nonumber \\
  C(2,3) &=&-s_\beta^2(c_\alpha^2-s_\beta^2), \nonumber \\
   C(3,2) &=& +{1\over 2}[c_\beta^2(2s_\beta^2-1) +c_\alpha^2
            +2s_\alpha c_\alpha s_\beta c_\beta], \nonumber \\
  C(3,3) &=& -{1\over 2}[c_\beta^2(2s_\beta^2-1)+s_\alpha^2
            +2s_\alpha c_\alpha s_\beta c_\beta], \nonumber \\
  C(4,2) &=& +{1\over 2}[s_\beta^2(2c_\beta^2-1) +s_\alpha^2
            +2s_\alpha c_\alpha s_\beta c_\beta].  \nonumber \\
  C(4,3) &=& -{1\over 2}[s_\beta^2(2c_\beta^2-1)+c_\alpha^2
            +2s_\alpha c_\alpha s_\beta c_\beta].  
\eqa
and 
\bqa
    \delta X &=& Re [\Pi_{AA}^\prime (0) 
         +2c_Ws_W {{\Pi_{AZ}(0)}\over{M_W^2}}  \nonumber \\
     && ~~    + {{(c_W^2-s_W^2)\Pi_W(M_W^2)-c_W^4\Pi_{ZZ}(M_Z^2)}\over
            {s_W^2M_W^2}}~] , \\
    \delta Y &=& -{1\over{M_{H^0}^2}}\Bigl[
       Re\Sigma_{H^0H^0}(M_{H^0}^2)
      +\sin^2(\alpha-\beta)\delta M_A^2
      +\cos^2(\alpha+\beta)M_Z^2\delta Z_Z \nonumber \\
     &&~~~~~~~~
        + {g\over{2M_W}} \cos(\alpha-\beta)[1+\sin^2(\alpha-\beta)]
                       \delta T_{H^0} \nonumber \\
     &&~~~~~~~~ 
       + {g\over{2M_W}} \sin(\alpha-\beta)\cos^2(\alpha-\beta)
                       \delta T_{h^0}\Bigr] , 
\eqa
with
\bq
     \delta Z_Z = 2c_W^2\delta Z_g-3c_W^2\delta Z_W-s_W^2\delta Z_B.
\eq
The tadpole counterterms for the physical Higgs bosons are defined by
\bq
   \llgm{ \delta T_{H^0} \cr \delta T_{h^0}  }\rrgm =
    \llgm{ \cos\alpha & \sin\alpha \cr 
          -\sin\alpha & \cos\alpha \cr}\rrgm
    \llgm{\delta {\cal T}_1 \cr \delta {\cal T}_2  } \rrgm .
\label{a.22}
\eq
The counterterm of $\tan\beta$ is then determined by 
\bq
     \delta\tan\beta  
       =- {1\over 2}\tan\beta(\delta Z_{H_1}-\delta Z_{H_2}
           -2{{\delta v_1}\over {v_1}}+2 {{\delta v_2}\over {v_2}}), 
\eq
while the gauge-boson mass counterterms are given by
\bqa
    \delta M_W^2 &=&M_W^2\lbrack
        2\delta Z_g -3\delta Z_W + \cos^2\beta \delta Z_{H_1}
                                 + \sin^2\beta \delta Z_{H_2} \nonumber \\
     & & ~~~~~~-2\cos^2\beta {{\delta v_1}\over{v_1}} 
              -2\sin^2\beta {{\delta v_2}\over{v_2}}\rbrack,\\ 
    \delta M_Z^2 &=&M_Z^2\lbrack  \delta Z_Z 
         + \cos^2\beta \delta Z_{H_1} + \sin^2\beta\delta Z_{H_2} \nonumber \\
     & &~~~~~~ -2\cos^2\beta {{\delta v_1}\over{v_1}} 
              -2\sin^2\beta {{\delta v_2}\over{v_2}}\rbrack,
\eqa
\noindent $\underline{\rm Chargino~ sector}$\par
\bqa
   \llgm{ \delta Z_{\lambda^w} \cr \delta Z_{\tilde H_1} \cr 
              \delta Z_{\tilde H_2}}\rrgm
  &=& \llgm{  1 & (s_L^2c_R^2+c_L^2s_R^2)/X & -2s_L^2s_R^2/X \cr
            1 & (2s_L^2c_R^2-c_L^2-c_R^2)/X & 2s_L^2c_R^2/X \cr
            1 & (2c_L^2s_R^2-c_L^2-c_R^2)/X & 2c_L^2s_R^2/X}\rrgm
   \llgm{ A \cr Re\Sigma_{11}^A(m_{\tilde\chi_1^+}) \cr 
            Re\Sigma_{22}^A(m_{\tilde\chi_2^+})}
    \rrgm , \nonumber \\
    \\
   \delta M_2 &=& {{s_Ls_R(\Delta_{22}-\epsilon_L \Pi_{22})
                   -c_Lc_R(\Delta_{11}-\Pi_{11})}\over
                   {c_L^2-s_R^2}} , \\
   \delta \mu &=& {{s_Ls_R(\Delta_{11}-\Pi_{11})
                   -c_Lc_R(\Delta_{22}-\epsilon_L \Pi_{22})}\over
                   {c_L^2-s_R^2}} , 
\eqa
where
\bqa
   X&=& s_L^2-s_R^2,  \\
   A&=& 2m_{\tilde\chi_1^+}[Re\Sigma^{S\prime}_{11}(m_{\tilde\chi_1})
              + m_{\tilde\chi_1^+}Re\Sigma^{V\prime}_{11}(m_{\tilde\chi_1^+})]
              +Re\Sigma^V_{11}(m_{\tilde\chi_1^+}), \\
    \Delta_{11}&=&
       + (c_Lc_R M_2 + c_Ls_R M_W{{\cos\beta}\over{\sqrt 2}}
                      + s_Lc_R M_W{{\sin\beta}\over{\sqrt 2}})
                      \delta Z_{\lambda^w} \nonumber \\
     && +(c_Ls_R \sqrt 2 M_W \cos\beta + s_Ls_R \mu) {1\over 2}
                      \delta Z_{\tilde H_1}\nonumber \\
     && +(s_Lc_R \sqrt 2 M_W \sin\beta + s_Ls_R \mu) {1\over 2}
                      \delta Z_{\tilde H_2}\nonumber \\
     && +c_Ls_R \sqrt 2 M_W \cos\beta ({1\over 2}\delta Z_{H_1}
                      + {{\delta g}\over g}
                     -{{\delta v_1}\over {v_1}})\nonumber \\
     && + s_Lc_R \sqrt 2 M_W \sin\beta ({1\over 2}\delta Z_{H_2}
                      + {{\delta g}\over g}
                        -{{\delta v_2}\over {v_2}}), \\
    \Delta_{22}&=&
      + (s_Ls_R M_2 - s_Lc_R M_W{{\cos\beta}\over{\sqrt 2}}
                      - c_Ls_R M_W{{\sin\beta}\over{\sqrt 2}})
                      \delta Z_{\lambda^w} \nonumber \\
     && +(-s_Lc_R \sqrt 2 M_W \cos\beta + c_Lc_R \mu) {1\over 2}
                      \delta Z_{\tilde H_1}\nonumber \\
     && +(-c_Ls_R \sqrt 2 M_W \sin\beta + c_Lc_R \mu) {1\over 2}
                      \delta Z_{\tilde H_2}\nonumber \\
     && -s_Lc_R \sqrt 2 M_W \cos\beta ({1\over 2}\delta Z_{H_1}
                      + {{\delta g}\over g}
                      -{{\delta v_1}\over {v_1}})\nonumber \\
     && -c_Ls_R \sqrt 2 M_W \sin\beta ({1\over 2}\delta Z_{H_2}
                      + {{\delta g}\over g}
                        -{{\delta v_2}\over {v_2}}),\\
 \Pi_{11}&=& 2m^2_{\tilde\chi_1^+}[Re\Sigma_{11}^{S\prime}(m_{\tilde\chi_1^+})
          +m_{\tilde\chi_1^+} Re\Sigma_{11}^{V\prime}(m_{\tilde\chi_1^+})]
                -Re\Sigma_{11}^S(m_{\tilde\chi_1^+}),  \\
 \Pi_{22}&=& {{m_{\tilde\chi_2^+}}\over 2}[(s_L^2+s_R^2)\delta Z_{\lambda^w}
         + c_R^2\delta Z_{\tilde H_1}+ c_L^2\delta Z_{\tilde H_2}] \cr
         &&  -Re\Sigma_{22}^S(m_{\tilde\chi_2^+})
            -m_{\tilde\chi_2^+}Re\Sigma_{22}^V(m_{\tilde\chi_2^+}).
\eqa

\noindent $\underline{\rm Neutralino~ sector}$\par
\bqa
   \delta Z_\lambda &=& {1\over{{\cal O}_{11}^2}}
         \bigl[\delta Z_{11}-{\cal O}_{12}^2 \delta Z_{\lambda^w}
         -{\cal O}_{13}^2 \delta Z_{\tilde H_1}
         -{\cal O}_{14}^2 \delta Z_{\tilde H_2}\bigr] , \\
   \delta M_1 &=& {1\over{{\cal O}_{11}^2}}
        \Bigl[ {{\delta m_{11}}\over{\eta_1^{*2}}} -\sum_{(p,q)\ne(1,1)}
        {\cal O}_{1p}{\cal O}_{1q}(\delta M_N)_{pq}
       -\sum_{(p,q)}{\cal O}_{1p}{\cal O}_{1q}\delta Z_p
        (M_N)_{pq} \Bigr] , \nonumber \\
\label{a.36}
\eqa
where ${\cal O}$ is the orthogonal matrix which diagonalizes
the neutralino mass matrix $M_N$
\bq
    {\cal O}M_N{\cal O}^t = \llgm{ m_{\tilde n_1} & & & \cr
                                   & m_{\tilde n_2} & & \cr
                                   & & m_{\tilde n_3} & \cr
                                   & & & m_{\tilde n_4} \cr}\rrgm
\label{a.37}
\eq
and
\bq
  \delta M_N=
     \llgm{\delta M_1 & 0   & -M_Z~s_W \cos\beta \Delta_{13} &
                              ~M_Z~s_W \sin\beta \Delta_{14}\cr
           *   & \delta M_2 & ~M_Z~c_W \cos\beta \Delta_{23} &
                              -M_Z~c_W \sin\beta \Delta_{24} \cr
           *   & *   & 0                & -\delta \mu        \cr
           *   & *   & *                & 0           \cr}\rrgm ,
\eq
with
\bqa
   \Delta_{13} &=& {{\delta M_Z^2}\over{2M_Z^2}}
                 -c_W^2\bigl({{\delta g}\over g}
                            -{{\delta g^\prime}\over {g^\prime}})
                 -\tan\beta\cos^2\beta\delta\tan\beta,  \\
   \Delta_{14} &=&{{\delta M_Z^2}\over{2M_Z^2}}
                 -c_W^2\bigl({{\delta g}\over g}
                            -{{\delta g^\prime}\over {g^\prime}})
                 +\cot\beta\cos^2\beta\delta\tan\beta,\\
   \Delta_{23} &=& {{\delta M_Z^2}\over{2M_Z^2}}
                 +s_W^2\bigl({{\delta g}\over g}
                            -{{\delta g^\prime}\over {g^\prime}})
                 -\tan\beta\cos^2\beta\delta\tan\beta, \\
   \Delta_{24} &=& {{\delta M_Z^2}\over{2M_Z^2}}
                 +s_W^2\bigl({{\delta g}\over g}
                            -{{\delta g^\prime}\over {g^\prime}})
                 +\cot\beta\cos^2\beta\delta\tan\beta.
\eqa
The phase factor $\eta_i$ in (\ref{a.36}) is 
to convert  the negative mass eigenvalue in (\ref{a.37}) to positive,
\bq
   \eta_i =\cases{ 1 &$m_{\tilde n_i}>0$ \cr
                   i &$m_{\tilde n_i}<0$}.
\eq 

\noindent $\underline{\rm sfermion ~sector}$\\
For simplicity, we show only the expression for the first generation.
\bqa
     \delta m_{\tilde f}^2 &=& -Re \Sigma_{\tilde f\tilde f}(m_{\tilde f}^2),~~~ 
       f= u_1, u_2, d_1, d_2, e_1,e_2,\nu_e,   \\
     \delta Z_{\tilde f\tilde f} &=& \Sigma^\prime_{\tilde f\tilde f}
          (m_{\tilde f}^2),~~~~~~~~~
       f= u_1, u_2, d_1, d_2, e_1,e_2,\nu_e,  \\
     {1\over 2}\delta Z_{\tilde f_i\tilde f_j} &=& 
       -{{\Sigma_{\tilde f_i\tilde f_j}
          (m_{\tilde f_j}^2)}\over{ m_{\tilde f_i}^2-m_{\tilde f_j}^2}},~~~ 
       i \ne j,~~f= u, d, e.    \\
    \delta\theta_e &=& {{ \delta m_{\tilde \nu_e} -\delta(M_W^2\cos2\beta-m_e^2)
                         -\cos^2\theta_e\delta m^2_{\tilde e_1}
                         -\sin^2\theta_e\delta m^2_{\tilde e_2} }\over
                         {\sin2\theta_e (m^2_{\tilde e_2}-m^2_{\tilde e_1}) }}, \nonumber \\
   \\
    \delta\theta_u &=& {1\over 2}{{\Sigma_{\tilde u_1\tilde u_2}(m_{\tilde u_1}^2)
                     +\Sigma_{\tilde u_1\tilde u_2}(m_{\tilde u_2}^2) }\over
                     {m_{\tilde u_2}^2-m_{\tilde u_1}^2}}
\eqa
{\footnotesize{
\bqa
    \delta\theta_d = {{ \delta(\cos^2\theta_u m^2_{\tilde u_1} +\sin^2\theta_u m^2_{\tilde u_2}
                          -M_W^2\cos2\beta-m_u^2+m_d^2)
                         -\cos^2\theta_d \delta m^2_{\tilde d_1}
                         -\sin^2\theta_d \delta m^2_{\tilde d_2}  }\over
                         {\sin2\theta_d (m^2_{\tilde d_2}-m^2_{\tilde d_1}) }},
\nonumber
\eqa
}}
\bqa
~ \label{a.49}
\eqa

\noindent $\underline{\rm QCD~ sector}$\par
\bqa
     \delta M_3 &=& -m_{\tilde g} [ Re\Sigma^S_{\tilde g}(m_{\tilde g}) 
                 + Re\Sigma^V_{\tilde g}(m_{\tilde g})] ,\\
     \delta Z_{\tilde g} &=& Re\Sigma^V_{\tilde g}(m_{\tilde g}) +
             2m_{\tilde g}^2 [ Re\Sigma^{S\prime}_{\tilde g}(m_{\tilde g})
                    + Re\Sigma^{V\prime}_{\tilde g}(m_{\tilde g})],\\
      \delta Z_{\rm gluon} &=&Re\Pi^{\prime}_{gg}(0), \\
    \delta Z_{g_s} &=&- C[ {2\over{4-d}}-\gamma_E+ln(4\pi)].
\eqa
where $C$ is the finite constant appearing at the one-loop vertex correction as
\bq
     (iV_{\mu\nu\lambda}^{abc})  [C({2\over{4-d}}-\gamma_E+ln(4\pi))
                + \cdots].  
\eq

\setcounter{section}{2}
\setcounter{equation}{0}
\section*{Appendix B.  External wavefunction renormalization constants}
We list the external wavefunction renormalization constant $\delta Z^{ext}$
which appears in the amplitude as
\bq
   {\cal M}\sim   {1\over 2}\delta Z^{ext}\times ({\rm Born~amplitude})
\eq
\noindent $\underline{\rm Gauge~bosons}$\\
\bqa
  \delta Z^{\rm ext}_W&=& \hat\Pi^\prime_W(M_W^2)
                       =\Pi_W^\prime(M_W^2)-\delta Z_W,    \label{b.2}\\
  \delta Z^{\rm ext}_Z&=& \hat\Pi^\prime_{ZZ}(M_Z^2)
                       =\Pi_{ZZ}^\prime(M_Z^2)-\delta Z_{ZZ}, \label{b.3} 
\eqa
\noindent $\underline{\rm Higgs~sector}$ \\
\bq
    \delta Z^{\rm ext}_{H^0} =\hat\Sigma^\prime_{H^0H^0}
           (M_{H^0}^2).
\label{b.4}
\eq
Since  the pole mass  of $h^0$ and $H^\pm$ in one-loop order 
does not agree with the tree mass, 
some complication appears.
\bq
    \delta Z^{\rm ext}_{h^0} =
    {1\over{1-\hat\Sigma_{hh}^{*\prime}(M_{h^0}^2, m_{h^0}^2,m_{H^0}^2)}}
    -1
   \approx \hat\Sigma_{hh}^{*\prime}(M_{h^0}^2, m_{h^0}^2,m_{H^0}^2)
\label{b.5}
\eq
where
\bqa
   \hat\Sigma_{hh}^{*\prime}(M_{h^0}^2)
   &\equiv&  {\partial\over{\partial q^2}}
    \hat\Sigma_{hh}(q^2)\Big|_{M^2_{h^0}} 
      \label{b.6}\\
   & &+{\partial\over{\partial q^2}}\Bigl[
    {{\hat\Sigma_{Hh}^2(q^2)}\over
     {q^2-m_{H^0}^2-\hat\Sigma_{HH}(q^2) }}\Bigr]
    \Big|_{M_{h^0}^2}. \nonumber 
\eqa
and $M_{h^0}$ is the one-loop improved pole mass of $h^0$.
The renormalized selfenergy functions are defined by
\bqa
  \hat\Sigma(q^2)_{hh} = \Sigma(q^2)_{hh}
     + \delta M^2_{hh}
     -q^2(\sin^2\alpha Re\delta Z_{H_1} + \cos^2\alpha Re\delta Z_{H_2}) ,~~~
        \\ 
 \hat\Sigma(q^2)_{HH} = \Sigma(q^2)_{HH}
     + \delta M^2_{HH}
     -q^2(\cos^2\alpha Re\delta Z_{H_1} + \sin^2\alpha Re\delta Z_{H_2}) ,
        \\ 
 \hat\Sigma(q^2)_{Hh} = \Sigma(q^2)_{Hh}
     + \delta M^2_{Hh}
     -q^2\cos\alpha \sin\alpha( Re\delta Z_{H_2} - Re\delta Z_{H_1}) ,~~~
\eqa
with
\bqa
    \delta M^2_{hh} &=& \cos^2(\alpha-\beta) \delta M_A^2 \label{b.10} \\
      & &-    {g\over{2M_W}} \cos(\alpha-\beta)\sin^2(\alpha-\beta)
                           T_{H^0}^{\rm loop} \nonumber \\
      & &- {g\over{2M_W}} \sin(\alpha-\beta)[1+\cos^2(\alpha-\beta)]
                           T_{h^0}^{\rm loop} \nonumber \\
      & &+    M_Z^2\lbrack \sin^2(\alpha+\beta)
               ( \delta Z_Z +\delta Z_{H_1} + \delta Z_{H_2}) \nonumber \\
      & & ~~~~~~~
            + \sin(\alpha+\beta)\sin(\alpha-\beta)
              (\delta Z_{H_1} - \delta Z_{H_2} ) \rbrack  \nonumber \\
     & &+ 2\sin(\alpha+\beta)\bigl[\sin\beta\cos\beta\cos(\alpha+\beta)
              -\sin\alpha\cos\beta\bigr] M_Z^2
                {{\delta v_1}\over {v_1}} \nonumber \\
     & &-2\sin(\alpha+\beta) \bigl[\sin\beta\cos\beta\cos(\alpha+\beta)
              +\cos\alpha\sin\beta\bigr] M_Z^2
                {{\delta v_2}\over {v_2}},  \nonumber \\
    \delta M^2_{HH} &=&- Re \Sigma_{HH}(M_H^2)
      +M^2_{H^0}(\cos^2\alpha\delta Z_{H_1}+\sin^2\alpha\delta Z_{H_2}),\\
    \delta M^2_{Hh} &=&-\sin(\beta-\alpha)
                           \cos(\beta-\alpha)\delta M_A^2 \label{b.12} \\
     & &+     {g\over{2M_W}} \sin^3(\beta-\alpha) T_{H^0}^{\rm loop}
           + {g\over{2M_W}} \cos^3(\alpha-\beta) T_{h^0}^{\rm loop} \nonumber \\
     & &-     M_Z^2\lbrack \sin(\alpha+\beta)\cos(\alpha+\beta)
               ( \delta Z_Z +\delta Z_{H_1} + \delta Z_{H_2}) \nonumber \\
     & &~~~~~~~
            + \sin\alpha\cos\alpha
               (\delta Z_{H_1} - \delta Z_{H_2} ) \rbrack \cr
     & &+ \bigl[{1\over 2}\sin(2\alpha+2\beta)(1+\cos 2\beta)
              -{{\sin 2\alpha\cos 2\alpha\sin 2\beta}\over
                {\sin(2\alpha-2\beta)}} \bigr]M_Z^2
                {{\delta v_1}\over {v_1}} \nonumber \\
     & &+ \bigl[{1\over 2}\sin(2\alpha+2\beta)(1-\cos 2\beta)
              +{{\sin 2\alpha\cos 2\alpha\sin 2\beta}\over
                {\sin(2\alpha-2\beta)}} \bigr]M_Z^2
                {{\delta v_2}\over {v_2}}.  \nonumber 
\eqa
The expressions (\ref{b.5}) and (\ref{b.6}) agree with those 
given in \cite{heinemeyer}.
\bq
   \delta Z^{\rm ext}_{H^\pm}
   = {1\over{1-\hat\Sigma_{H^\pm H^\pm}^{*\prime}(M_{H^\pm}^2)}}-1 
     \sim \hat\Sigma_{H^\pm H^\pm}^{*\prime}(M_{H^\pm}^2), 
\label{b.13} 
\eq
where
\bqa
   \hat\Sigma_{H^\pm H^\pm}^{*\prime}(q^2)
   &=&  {\partial\over{\partial q^2}}
    \hat\Sigma_{H\pm H^\pm}(q^2)\Big|_{M^2_{H^\pm}} 
      \label{b.14}\\
   & &+{\partial\over{\partial q^2}}\Bigl[
    {{\hat\Sigma_{H^\pm G^\pm}^2(q^2)}\over
     {q^2-m_{H^\pm}^2-\hat\Sigma_{H^\pm H^\pm}(q^2) }}\Bigr]
    \Big|_{M_{H^\pm}^2}. \nonumber 
\eqa
The renormalized selfenergy functions appearing in (\ref{b.14}) 
are given by
\bqa
 \hat\Sigma(q^2)_{H^\pm H^\pm} = \Sigma(q^2)_{H^\pm H^\pm}
   + \delta M^2_{H^\pm H^\pm}
   -q^2(s^2_\beta Re\delta Z_{H_1} + c^2_\beta Re\delta Z_{H_2}) , \\ 
 \hat\Sigma(q^2)_{H^\pm G^\pm} = \Sigma(q^2)_{H^\pm G^\pm}
   + \delta M^2_{H^\pm G^\pm}
   -q^2c_\beta s_\beta( Re\delta Z_{H_2} - Re\delta Z_{H_1}) ,~
\label{b.16}
\eqa
with
\bqa
    \delta M^2_{H^\pm H^\pm}&=& \delta M^2_{A^0 A^0}
       +M_W^2(\delta Z_x -2c^2_\beta {{\delta v_1}\over{v_1}} 
                         -2s^2_\beta {{\delta v_2}\over{v_2}}), \\ 
    \delta M^2_{H^\pm G^\pm}&=& \delta M^2_{G^0 A^0}
          -c_\beta s_\beta M_W^2({{\delta v_1}\over{v_1}}
                                -{{\delta v_2}\over{v_2}}).
\label{b.18}
\eqa
In the one-loop order, (\ref{b.5}) and (\ref{b.14}) become
\bqa
   \delta Z^{\rm ext}_{hh}&=&\hat\Sigma^\prime_{h^0h^0}(M^2_{h^0}) ,  
      \label{b.19}\\
   \delta Z^{\rm ext}_{H^\pm}&=&\hat\Sigma^\prime_{H\pm H^\pm}(M^2_{H^\pm}) .
      \label{b.20}
\eqa 
\noindent $\underline{\rm Chargino}$ \\
\bq
    \delta Z^{\rm ext}_{\tilde\chi^+_2}
   = 2M_{\tilde\chi^+_2}(\Sigma^{S\prime}_2(M_{\tilde\chi^+_2}^2)
                     +M_{\tilde\chi^+_2}\Sigma^{V\prime}_2(M_{\tilde\chi^+_2}^2))
         +\Sigma^V_2(M_{\tilde\chi^+_2}^2)-{1\over 2}(\delta Z^R_{22}+\delta Z_{22}^L) ,
    \label{b.21}
\eq
where the chargino selfenergy functions are decomposed as (\ref{a.14}).
In terms of the renormalization constants introduced in section 2,
the chargino wavefunction renormalization constants are given by
\bqa
     \delta Z^L_{22} &=& \sin\phi_L^2\delta Z_{\lambda^w} 
                       + \cos\phi_L^2\delta Z_{\tilde H_2} , \nonumber \\
     \delta Z^R_{22} &=& \sin\phi_R^2\delta Z_{\lambda^w} 
                       + \cos\phi_R^2\delta Z_{\tilde H_1} . 
\label{b.22}
\eqa

\noindent $\underline{\rm Neutralino}$\\
\bq
   \delta Z_{\tilde\chi^0_i}^{\rm ext} =
     2M_{\tilde\chi^0_i}[\Sigma^{S\prime}_{ii}(M^2_{\tilde\chi^0_i})
     +M_{\tilde\chi^0_i}\Sigma^{V\prime}_{ii}(M^2_{\tilde\chi^0_i})]
             +\Sigma_{ii}^V(M^2_{\tilde\chi^0_i})  -\delta Z_{ii} .
\label{b.23}
\eq
where $i=2,3,4$ and the selfenergy function of Majorana particles is
decomposed as
\bq
       \Sigma_f(\gslash q) \equiv \Sigma_f^S(q^2) {\bf 1} + 
       \Sigma_f^V(q^2)\gslash q .  
\eq
The neutralino wavefunction renormalization constant appearing in 
(\ref{b.23}) is expressed in terms of the basic renormalization 
constants introduced in section 2 as
\bq
   \delta Z_{ii} = \sum_k 
        (Re\delta Z)_k ({\cal O}_N)_{ik}({\cal O}_N)_{ik},~~~{\rm no~sum~over}~i ,
\eq
where 
\bq
   \delta Z_k \equiv (\delta Z_{\lambda}, \delta Z_{\lambda^w}, 
                    \delta Z_{\tilde H_1}, \delta Z_{\tilde H_2}) . 
\eq
and ${\cal O}$ is the orthogonal matrix which diagonalizes the neutralino
mass matrix. See (\ref{a.37}).

\vspace{0.2 cm}

Note that for unstable particles, even if the residue condition 
is imposed on the propagator  at the pole position, 
there is a non-vanishing $\delta Z^{ext}$ which is
ultraviolet-finite and purely imaginary. 
For example, we can easily check that $\delta Z_{\tilde\chi^+_1}^{\rm ext}$
which is obtained from (\ref{b.21}) by changing the index 2 to 1,
is purely imaginary.
We can neglect such  contributions if perturbation works.

\end{document}